\documentclass{aastex}          
\usepackage{spr-astr-addons}    

\begin{document}
%
\title{Dynamic screening in solar and stellar nuclear reactions}
 
\shorttitle{Dynamic screening in solar and stellar nuclear reactions}
\shortauthors{Mussack \& D\"appen}

\author{Katie Mussack} 
\affil{Institute of Astronomy, University of Cambridge, Cambridge, CB3
  0HA, UK}
\affil{Los Alamos National Laboratory, X-2, MS P365, Los Alamos, New
    Mexico, 87545, USA}
\and 
\author{Werner D\"appen}
\affil{Department of Physics and Astronomy, University of Southern
  California, Los Angeles, California, 90089, USA}


\begin{abstract}

In the hot, dense plasma of solar and stellar interiors, the Coulomb
interaction is screened by the surrounding plasma. Although the
standard Salpeter approximation for static screening is 
widely accepted and used in stellar modeling, the question of
dynamic screening has been revisited. In particular, Shaviv and Shaviv 
apply the techniques of molecular dynamics to the conditions in the
solar core in order to numerically determine the dynamic screening
effect. By directly calculating the motion of ions and electrons
due to Coulomb interactions, they compute the
effect of screening without the mean-field assumption inherent in the
Salpeter approximation. Here we reproduce their numerical analysis of
the screening energy in the plasma of the solar core and conclude that
the effects of dynamic screening are relevant and should be included
in the treatment of the plasma, especially in the computation of
stellar nuclear reaction rates. 

\end{abstract}


\section{INTRODUCTION}

Screening of nuclear reactions in stellar interiors is a hotly debated
issue.  The standard Salpeter treatment \citep{Salpeter_1954} was a
necessary improvement in early equation-of-state work. Many faithful
fans consider this as evidence that the 
traditional screening formulation must be used for solar
models. However, the incompatibility of models generated with the
recently revised solar abundances and helioseismic results highlights
the need to re-examine the physics used to develop and analyze solar
models \citep{Asplund_2009}. In this paper, we re-examine screening in
the solar core and present evidence that dynamic effects must be considered when
examining screening in stars.

\subsection{Electrostatic Screening}
\label{sect:static}

Under the extreme temperatures and densities of the solar core, the plasma 
is fully ionized. The free electrons and ions interact with a Coulomb 
potential energy
\begin{equation}
\label{eq1}
  U(r)=\frac{e^2}{r} \;.
\end{equation}
In this Coulomb system, 
nearby plasma is polarized by each ion. 
When two ions approach with the possibility of engaging in a 
nuclear reaction, each ion is surrounded by a screening cloud. Each 
ion is attracted to the electrons and repelled by the protons in its 
partner's cloud. The combined effect 
of the particles in the screening clouds on the potential energy of
the pair of ions is referred to as screening. This electrostatic
screening effect 
reduces the standard Coulomb potential between approaching ions to a screened potential which includes the contribution to the 
potential from the surrounding plasma. The reduced potential enables the ions
to tunnel through the potential barrier more easily, thus enhancing fusion
rates.  As illustrated by other authors at this meeting \citep{Dappen_2009,Baturin_2009,Straniero_2009,Yusof_2009}, understanding screening is important in solar and stellar modeling.

\citet{Salpeter_1954} derived an expression for the enhancement of 
nuclear reaction rates due to electron screening. By solving the 
Poisson-Boltzmann equation for electrons and ions in a plasma under the 
condition of weak screening ($\phi_{\rm{interaction}}<< k_BT$), Salpeter 
arrived at an expression for the screening energy that is equivalent to 
that of the Debye-H\"uckel theory of dilute solutions of electrolytes 
\citep{Debye_1923},
\begin{equation}
\label{eq2}
   U_{\rm{screen}} = \frac{e^2}{\lambda_D} 
\end{equation}
where the Debye length, $\lambda_D$, is the characteristic screening length of
a plasma at temperature $T$ with number density $n$, 
\begin{equation}
\label{eq3}
  \lambda_D^2 = \frac{\epsilon_0 k_B T}{ne^2}. 
\end{equation}

\subsection{Dynamic Screening}
\label{sect:dynamic}

Although Salpeter's approximation for screening is widely accepted, 
several papers over the last few decades  \citep[e.g.][]{Shaviv_1996, 
Carraro_1988, Weiss_2001} have questioned either the 
derivation itself or the validity of applying the approximation to hot, 
dense, Coulomb systems like the  plasma of the solar core. Various work 
deriving alternative formulae for \emph{electrostatic} screening
\citep{Carraro_1988, Opher_2000, Shaviv_1996, Savchenko_1999, Lavagno_2000,  
Tsytovich_2000} were refuted in subsequent papers \citep[see][for a summary of
arguments in Salpeter's defense]{Bahcall_2002}. However, the question of
\emph{dynamic} screening remains open. Dynamic screening
arises because the protons in a plasma are much slower than the electrons. They
are therefore not able to rearrange themselves as quickly around individual 
faster moving ions. Since nuclear reactions require energies several 
times the average thermal energy, the ions that are able to engage in 
nuclear reactions in the Sun are such faster moving ions, which
therefore may not be accompanied by their full screening cloud. Salpeter uses
the mean-field approach in which the many-body interactions are reduced to an
average interaction that simplifies calculations. This technique is quite
useful in calculations that rely on the average behavior of
the plasma. However, dynamic effects for the fast-moving, interacting ions lead
to a screened potential that deviates from the average value. The 
nuclear reaction rates will therefore differ from those computed with
the mean-field approximation. 

\citet{Shaviv_1996} used the method  of molecular 
dynamics to model the motion of charges in a plasma under solar 
conditions in order to investigate dynamic screening. The advantage of 
the molecular-dynamics method is that it does not assume a mean field. 
Nor does it assume a long-time average potential for the scattering
of any two charges, which is necessary in the statistical way to solve
Poisson's equation to obtain the mean potential in a plasma. Shaviv and 
Shaviv attribute the differences between their simulations and 
Salpeter's theory to dynamic effects. Since their claims have been met 
with skepticism, we have conducted independent molecular-dynamics 
simulations to confirm the existence of dynamic effects.

The viewpoint presented by Bahcall et al., 2002 can be summarized with
their statement ``There is only one right answer, but there are many
wrong answers.'' Although we agree that equation (\ref{eq2}) is
the right answer to the question of static screening, we contend that
this is not the right question to ask. All arguments in favor of
Salpeter's formulation rely on a mean-field treatment, an assumption
that must be tested before it is implemented. The work presented in
this paper addresses the 
more appropriate question ``is the mean-field approach applicable in
stellar plasma?'' Our work will show that there are deviations from
the mean field in the case of p-p reactions in the solar core and that
dynamic screening should be considered in order to obtain a more accurate
representation.

\section{METHOD}
\label{sect:method}

 How can we test the question of mean-field theory's relevance in
 solar plasma? \citet{Shaviv_1996} developed a
 method that relies on the techniques of molecular dynamics to model
 the behavior of solar plasma without using mean-field
 assumptions. Their simulations show deviations from mean-field
 theory that would lead to changes in nuclear reaction
 rate calculations. Their claims have been met with skepticism, so we
 replicated and analyzed their work in order to resolve the issue. In
 our previous work \citep{Mao_2004, Mao_2009, Mussack_2006, Mussack_2007}, we examined the methods and assumptions used in Shaviv and Shaviv's work, including their treatment of the long-range Coulomb force, the effective quantum potentials, and the system size. We did not find any problems with their techniques. Furthermore, we confirmed that the mean-field theory does not adequately describe the behavior of the plasma.

Here we show that the screening energy of two interacting protons in our simulation depends on the relative kinetic energy of the pair. We also determine the dynamically screened interaction potential energy and discuss the significance of this result. 
 
In order to numerically determine the effect of dynamic screening on
p-p reactions, we modeled a 3-dimensional box of 500 protons and 500 electrons
interacting via the Coulomb potential. The effective pair
potentials derived for a hydrogen plasma by \cite{Barker_1971,
  Deutsch_1977, Deutsch_1978, Deutsch_1979} were employed to include quantum
corrections. The temperature and density of
the solar core ($T=1.6 \rm{x} 10^7 \;\rm{K}$, $\rho = 1.6 \rm{x} 10^5
\; \rm{kg/m^3}$)  were used to determine the velocities and density of
the particles in the box. A thermostat was implemented to maintain
constant temperature. Periodic boundary conditions and the
minimum-image convention were applied. The velocity verlet algorithm followed
the time evolution of the system. See \cite{Mao_2009} for more details.

The screening energy was calculated for pairs of approaching
protons in the following way. For each proton, we designated the
nearest approaching proton 
as its partner. Then we tracked each pair of protons through their approach and
subsequent retreat. At the point of closest approach, we recorded the
separation, $r_c$, and the kinetic energy of the pair,
$E_{kinetic}(r_c)$. When the pair was separated by a sufficiently great
distance, $R_f$, we recorded the kinetic energy of the pair,
$E_{kinetic}(R_f)$, and stopped tracking that pair (for our simulations,
$R_f = 2 \; a_B$ where $a_B$ is the Bohr radius). At this point, we
designated a new partner and repeated the tracking process. 

The screening energy of each pair was computed from the difference in
energy at $r_c$ and at $R_f$
\begin{equation}
\label{eq4}
  E_{\rm{screen}} =E_{\rm{pair}}(r_c) - E_{\rm{pair}}(R_f) .
\end{equation}
This represents the energy exchanged between a pair and the surrounding
plasma. Equation \ref{eq4} can be expanded as 
\begin{equation}
\label{eq5}
  E_{\rm{screen}}= \left( E_{\rm{kinetic}}(r_c) + \frac{e^2}{r_c} \right) 
               - \left( E_{\rm{kinetic}}(R_f) + \frac{e^2}{R_f} \right).
\end{equation}

\section{RESULTS}
\label{sect:results}

\begin{figure}
\includegraphics[width=82mm]{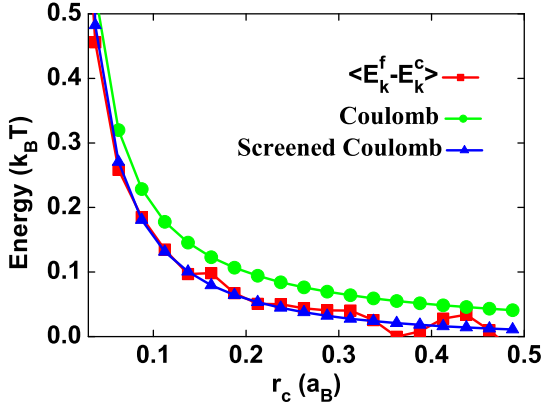}
\vskip -0.4cm \caption{Dependence of the average kinetic energy change
$<\!\!\triangle E_{\rm{kinetic}}\!\!>$ on the closest distance $r_c$, compared
with the Coulomb potential and screened Coulomb potential.} 
\label{plot1} \vskip 0.3cm
\end{figure}

\begin{figure}
\includegraphics[width=82mm]{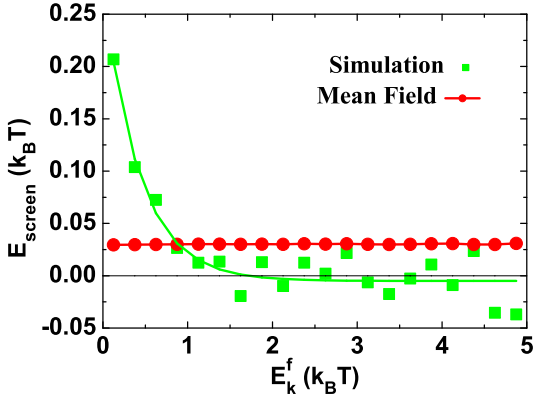}
\vskip -0.4cm \caption{Dependence of the screening energy from the
  simulation (squares) on the far-apart kinetic energy
$E_{kinetic}^f$. The Debye-H\"uckel screening energy computed at the averege
closest-approach distance of pairs of protons with a given far-apart
kinetic energy is shown (circles) for comparison.}
\label{plot2} \vskip .3cm
\end{figure}

A key ingredient for the screening energy in equation \ref{eq5} is the
difference in the kinetic energy of a pair when partner protons are far
apart and when they are at their closest separation.
\begin{equation}
\label{eq6}
  \Delta E_{\rm{kinetic}} = E_{\rm{kinetic}}(R_f) - E_{\rm{kinetic}}(r_c) .
\end{equation}
In figure \ref{plot1}, we show the average change in kinetic energy of
approaching pairs for each distance of closest approach. For
comparison, we also plot the bare Coulomb potential and the statically
screened Coulomb potential as a function of separation.
We see that at the distance of closest approach, the average kinetic
energy exchanged between a pair of protons and the plasma closely
matches the statically screened potential. This confirms that
Salpeter's static screening can successfully describe \emph{average} properties
of the system. However, we can see the dynamic effect on screening when we
sort the pairs of particles by relative kinetic energy. Figure
\ref{plot2} shows the average energy gained from the plasma by pairs
of protons with a given far-apart kinetic energy. This is quite
different from the Debye-H\"uckel screening energy calculated for the
average closest-approach distance of pairs in each kinetic energy bin.
We see that pairs of protons with a kinetic energy less than the
thermal energy of the plasma gain more energy from the surrounding
plasma than the mean-field result, while pairs with a kinetic energy
greater than the thermal energy gain less energy than the mean-field
result and even tend to lose energy to the plasma. From this plot, we
can estimate the screening
energy of the p-p reaction at the Gamow energy of $4.8 \; \rm{kT}$. 

In order to quantify the dynamic effect on screening in the plasma of
the solar core, we have split the total interaction energy into the Coulomb
contribution and the interaction energy from the plasma,
\begin{equation}
\label{eq7}
  U(r) = \frac{e^2}{r} - E_{\rm{screen}}(r) .
\end{equation}
For the mean-field treatment,
\begin{equation}
\label{eq8}
  E_{\rm{screen,mf}}(r) = \frac{e^2}{r} \left( 1- {\rm{exp}}
  \left(-r / \lambda_D \right) \right).
\end{equation}

Because there is no formalism to compute the dynamic effect
analytically, we use the simulation results to determine $E_{\rm{screen}}$ for
dynamic screening. One key difference from the static screening
expression is that the dynamic screening energy is a function of both
pair separation and relative velocity. As before, we split the total
interaction energy into the Coulomb and screening cloud contributions. 
\begin{equation}
\label{eq9}
   U(r,v) = U_{\rm{Coulomb}}(r) - E_{\rm{screen,dyn}}(r,v)
\end{equation}

\begin{figure}
  \includegraphics[width=82mm]{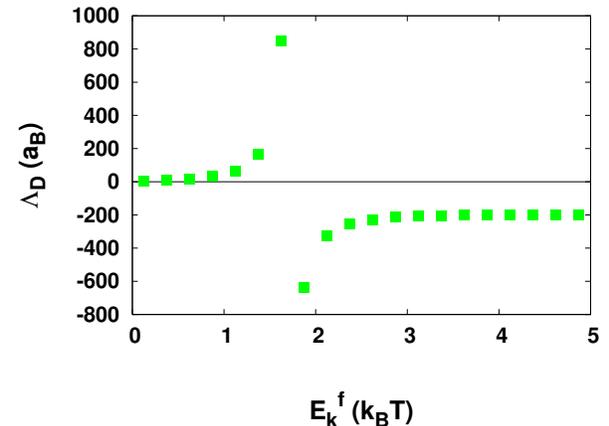}
 \vskip -0.4cm \caption{Modified Debye length for dynamic
   screening at the distance of closest approach obtained from simulations.}
\label{plot3} \vskip .3cm
\end{figure}

For comparison, the dynamic screening energy at the distance of
closest approach can be written in a form similar to the static screening energy 
\begin{equation} 
\label{eq10}
   E_{\rm{screen,dyn}}(r_c,v) = \frac{e^2}{r_c} \left(
      1- {\rm{exp}}(-r_c/{\Lambda_D(v)}) \right)
\end{equation}

by including a new velocity dependent Debye-like radius,
$\Lambda_D(v)$. Figure \ref{plot3} shows the form of $\Lambda_D(v)$
determined from the simulations. 

\section{DISCUSSION}
\label{sect:discussion}

This work confirms that screening in the hot, dense plasma of stellar
cores depends on the relative velocities of the interacting ions. The
Debye-H\"uckel screening energy is only valid for describing average
properties of the plasma. Since faster ions are more likely to engage
in nuclear reactions than thermal ions, the mean-field treatment does
not provide an accurate representation of this velocity-skewed
phenomenon. In fact, the fast pairs of ions tend to lose energy to the
plasma instead of gaining energy from it which would reduce nuclear
reaction rates instead of enhancing them. Solar and stellar models
should be adjusted to account for this dynamic screening effect. 

Currently, there is no formalism to compute dynamic
screening analytically. This paper is intended to provide
insight into the difference between the Salpeter formalism and the
numerically determined dynamic screening. A detailed calculation
of the dynamic screening correction to the p-p reaction rate in the
solar core based on this numerical work is underway for a future
publication. However, these numerical calculations will not minimize the need
for an analytical formalism for dynamic screening in order to
generalize the results to other temperatures, densities, compositions,
and reactions. Only then can dynamic screening be encorporated
consistently in solar and stellar models.

 \acknowledgments

We thank Dan Mao for helpful discussions about the simulations. This
work was supported in part by grant AST-0708568 of the National
Science Foundation.


%

%

\end{document}